\documentstyle[epsfig,psfig,epsf,aps,prl,multicol]{revtex}
\begin{document}
\draft
\title{Bandgap renormalization and excitonic binding in
T-shaped quantum wires}
\author{M. Stopa} 
\address{
Walter Schottky Institute \\ 1 Am Coulombwall \\ D-85748 Garching, Germany \\
phone 49-89-289-12198, Fax: 49-89-289-12737 \\ e-mail: stopa@wsi.tum.de}
\date{\today}
\maketitle

\begin{multicols}{2}
[
\begin{abstract}
We calculate the electronic structure for a modulation doped and
gated T-shaped quantum wire using density functional theory.
We calculate the bandgap renormalization as a function of
the density of conduction band electrons, induced by
the donor layer and/or the gate, for the translationally
invariant wire, incorporating all growth and geometric
properties of the structure completely. We show that most of
the bandgap renormalization arises from exchange-correlation
effects, but that a small shift also results from the difference
of wave function evolution between electrons and holes.
We calculate the binding energy of excitons in the wire, which
breaks translational invariance, using a simpler, cylindrical
model of the wire. For a single hole and a one dimensional 
electron gas of density $n_e$, screening of the exciton binding
energy is shown to approximately compensate for bandgap
renormalization, suggesting that the recombination energy
remains approximately constant with $n_e$, in agreement with
experiment. We find that the nature of screening, as treated
within our non-linear model, is significantly different from
that of the various linear screening treatments, and the
orthogonality of free carrier states with the bound electron
states has a profound effect on the screening charge. We find that
the electron and hole remain bound for all densities up
to $\sim 3 \times 10^6 \; cm^{-1}$ and that, as $n_e$ increases
from zero, trion and even ``quadron'' formation becomes allowed.
\end{abstract}
]

\section{Introduction}

The gap in a semiconductor heterostructure between conduction
and valence bands, and the interaction between electrons in the
one and holes in the other, are known to depend in a complicated
fashion on the presence of mobile charges \cite{Glu97,Ros96,Bri97}. 
The interpretation of optical experiments in doped semiconductor 
quantum wires \cite{Weg93,Amb97},
for example, must invoke the variation of band edges due to
many-body and geometrical effects, the ``red shift'' associated with
electron-hole binding into excitons and finally the reduction of
the exciton binding energy due to screening by free carriers.

Theoretical description of bandgap renormalization (BGR) and
exciton formation and screening are frequently addressed with
a many-body formalism \cite{Zim78,Das99}. Within this framework,
contribution of both electron-phonon and electron-electron
self-energies to BGR \cite{Hwa98} as well as the influence of
dynamical screening on the exciton binding energy can be
studied. The many-body treatment, however, has the disadvantage
that, for BGR it commonly ignores geometrical factors, such as the
quantum confined Stark effect \cite{Blo89}, whose relevance is
structure-specific. Furthermore, in the exciton problem, many-body
theory treats screening within the linear approximation and,
generally, influence of the bound electron on the free electrons
is not fully included. In particular the orthogonality of the
free electron states with the bound state, which increases in
importance in lower dimensional systems, are typically not
included \cite{Gui85}.

A different formalism without these shortcomings, albeit one
which forsakes the electron-phonon interaction and the dynamics
of screening, is that of density functional theory (DFT) \cite{DFT}.
Within DFT the ground state energy of an interacting many particle
system is known rigorously to be given by a functional of 
the density. The essential problem of DFT is that this functional
is unknown and the components beyond kinetic and direct
electrostatic energy are isolated into an ``exchange-correlation''
functional which must be approximated radically. Typically the
limit wherein the density varies adiabatically, known as the
local density approximation (LDA), is assumed. Nonetheless, 
successful treatments of a wide class of systems; from
atoms and molecules to solids and heterostructures, abound
in the literature. It is the purpose of this paper to provide
a theoretical description of BGR and exciton screening, applied
particularly to semiconductor quantum wires (QWs), within DFT.

We are interested in quantitative comparison with optical 
experiments on T-shaped quantum wires (TQWs) in the presence
of a one-dimensional gas of electrons (1DEG) with 1D density $n_e$,
induced in
the conduction band via a combination of modulation doping and 
gating \cite{Sed99}. By way of warning, this one-component
plasma contrasts with
the case of many studies which focus on intrinsic QWs wherein
an overall charge neutral electron-hole plasma is generated 
entirely through photoexcitation. In our case photoexcitation
is assumed to provide a small number of holes (which we take 
as a single hole) and to have a negligible effect on the
density of conduction band electrons.

 \begin{figure}[hbt]
 \setlength{\parindent}{0.0in}
  \begin{minipage}{\linewidth}
 \epsfxsize=8cm
 \epsfbox{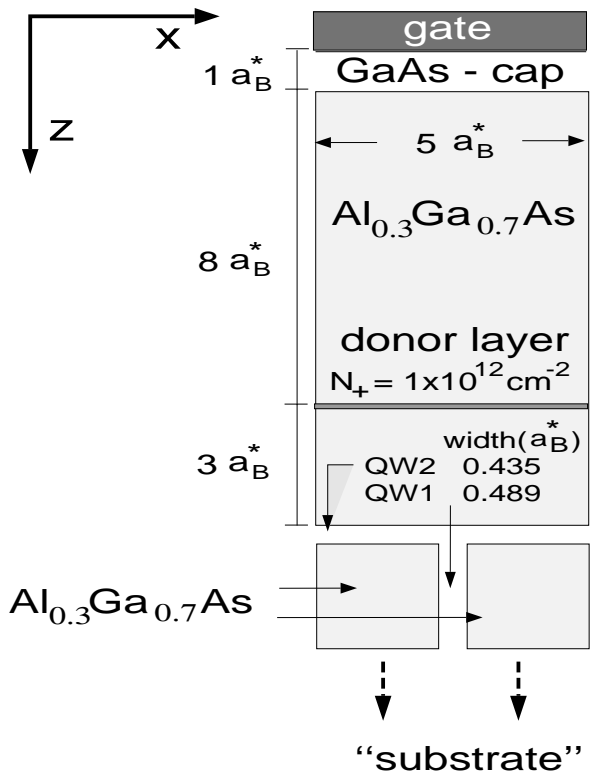}
 \vspace*{3mm}
 \caption{Schematic of T-shaped quantum wire cross section.}
  \end{minipage}
 \end{figure}

We evaluate the band structure of a TQW in two steps. In
the first step (section II)
we consider a realistic model of a 
cross-section of a TQW, shown in Fig. 1, and solve 
self-consistently for the electronic structure of
the subbands, as a function of the gate voltage,
assuming translational invariance along the wire.
Material-specific effective masses and band offsets, as
well as the actual growth profiles and doping densities 
of the structure are included. The exciton problem, which
breaks translation invariance, is treated in the second
step (section III) by a simplified, cylindrical model of a quantum wire
which employs the results of the first
step for its rationale and its parameters. Thus the philosophy
of the calculation is that the exciton formation and its
screening represent an additive correction to the 
translationally invariant band
structure of step 1. This exciton effect 
can be analyzed within a single subband,
generic wire model so long as the effects of lateral
confinement are correctly borrowed from the more faithful,
albeit 2D calculation in step 1.

Among our main results are the following. For the
translationally invariant band structure (section II): 
(1) We find a BGR as a function of $n_e$ with comparable order 
of magnitude to that derived within many-body theory. 
(2) One small ($\sim 20 \; \%$) component of this BGR arises from
a variation of the lowest electron subband eigenfunction 
with $n_e$. Specifically, at higher $n_e$ the lowest electronic
wave function spreads somewhat along the overgrowth
interface (throughout QW2 in Fig. 1) whereas it is more strongly confined
at the junction (i.e. the intersection of the two wells)
at lower $n_e$. The hole wave function, 
by contrast, is always localized at the junction.
(3) The stability of the hole eigenstate at the junction
is dependent on the boundary conditions used for large $z$
(see Fig. 1). If the chemical potential is pinned to an
assumed background donor density in the substrate 
(we use the term ``substrate'' to denote the region far from the
donor plane, i.e. large $z$, see Fig. 1), 
then the hole bound state becomes a
resonance which is metastable to
escape into the substrate. 
On the other hand, when the substrate is treated
as exactly electrically neutral (or pinned to acceptors) 
the hole state at the junction is stable. (4) For the specific 
dimensions and compositions which we consider \cite{Sed99}
the wires are predominantly one-dimensional with filling of no more than
two subbands and spreading of the density into
QW2 generally less than
$\sim 15 \; \%$ of the total density.

Regarding the exciton problem (section III), we find: (1) the 
exciton binding energy weakens considerably as the 1DEG density
is turned on, but never vanishes completely. 
 (2) The requirement of orthogonality between the 
bound electron eigenfunction and those of the screening 
electrons leads to an {\it anti-screening} of the exciton,
such that the density of free electrons in the vicinity
of the hole is depressed. This effectively positive charge
raises the energy of the hole while the energy of the electron
actually becomes more negative
and its localization about the hole increases with density.

Finally, the combination of the bandgap renormalization
and the exciton screening tend to cancel
and, consistent with experiments, the recombination
energy remains relatively constant with $n_e$. This result is plotted
in figure 6 but discussed at the end of section III.

We principally employ effective atomic
units wherein $1 \; Ry^* = m^* e^4/2 \hbar^2 \kappa^2  
\approx 5.25 \; meV$ and
$1 \; a_B^* = \hbar^2 \kappa/m^* e^2 \approx 100 $ \AA.
For comparison with experiments we also use $cm^{-1}$, where 
$1 \; a_B^{* -1} \approx 10^6 \; cm^{-1}$.

\section{Translationally invariant wire: BGR}

We calculate the electronic structure of the $GaAs-AlGaAs$
TQWs \cite{LP1}
by solving Schr\"{o}dinger and Poisson equations self-consistently,
within a region illustrated in Fig. 1,
for the conduction band electrons,
and by including exchange and correlation in the
local density approximation (LDA) \cite{DFT}.
There are numerous
calculations of BGR in lower dimensional systems which
employ many-body theory to determine the self-energy correction
to the subbands from exchange and correlation effects 
\cite{Das89,Sch89}. In DFT a bandgap renormalization
arises as follows. First, variations in the electrostatic potential
tend to affect both bands equally except for kinematic
effects like the quantum confined Stark effect \cite{Blo89}
where the electron
wave function is centered on a different spatial location
(at different potential) than that of the hole. By contrast,
exchange and correlation effects lower the energy of electrons
in proportion to $n_e$, the electron density,
and raise the energy of holes in proportion to $n_h$, the hole
density, producing an overall shrinkage of the bandgap.
For the one component plasma considered here, therefore,
the exchange-correlation potential acts only on the conduction
band electrons, since $n_h \approx 0$.

One advantage of our procedure is that while 
many-body calculations in principle begin from wave functions and
subband eigenenergies derived from some Hartree calculation,
the variation of these properties with density and the aforementioned
kinematical corrections to the bandgap are usually ignored. By
contrast DFT automatically includes these. Furthermore, the separate
contributions from kinematic effects as opposed to exchange and
correlation are easily isolated (to first order) by performing a 
calculation without the exchange-correlation potential (i.e. Hartree
only). 

We employ the parameterization of Ceperley and Alder \cite{Ceperley}
for the density-dependent, exchange-correlation 
potential $V_{xc}(n({\bf x}))$.
Taking $y$ as the translationally invariant direction (see figure 1),
the Schr\"{o}dinger equation in the $x-z$ plane ($k_y=0$) reads
\begin{eqnarray}
[-\nabla \frac{1}{2m^*(x,z)} \nabla & + & V_C (x,z) + 
e \phi (x,z) + \nonumber \\ 
V_{xc}(n(x,z))]  \psi_n(x,z) 
& = & E_e^n \psi_n(x,z) \label{eq:one}
\end{eqnarray}
where $V_C(x,z)$ is the conduction band offset \cite{Zachaw}
which depends only on the local aluminum concentration $\eta$
(we use the symbol $\eta$ instead of the standard
$x$ to avoid confusion with the coordinate $x$),
the 2D eigenfunctions and eigenvalues
are given by $\psi_n (x,z)$ and $E_n$ respectively,
and the electrostatic potential is $\phi(x,z)$. Also in equation 
\ref{eq:one} the level index is $n$, and the subscript
$e$ refers to electrons in the conduction band whereas for
holes we use $E_h^n$. The eigenstates are abbreviated
below as $en$ and $hn$ for electrons and holes, respectively. Periodic
boundary conditions are taken at $x = \pm a/2$ where
$a=5 \; a_B^*$ is the period of the TQW superlattice
as fabricated \cite{Sed99}. 
The effective mass dependence on position arises through its
dependence on $\eta$ \cite{Zachaw}. In the specific device we consider,
all regions have either $\eta=0$ (pure $GaAs$) or else
$\eta=0.3$ for the barrier regions. We ignore interface grading effects
as well as image effects, taking the dielectric constant as $\kappa=12.5$
everywhere.

For the solution of Poisson's Equ. we also assume periodic
boundary conditions at the $x$-borders. For the surface $z=0$
we simulate the surface metal gate in the standard fashion
\cite{LP1} by Dirichlet boundary conditions which fix
the gate potential modulo an offset of $0.8 \; eV$ for
the Schottky barrier. Variation of the electron sheet
density in the overgrowth well $N_e$ (or, equivalently,
the 1D density $n_e$) is accomplished by biasing
this surface gate. Other experimental methods for varying $n_e$ include
illumination or a
variation of the thickness of the spacer between the
modulation doping layer and the overgrowth well. While we have modeled
each of these methods independently, in fact in each case there is an 
approximately homogeneous sheet (or sheets) of positive charge 
balancing the electronic charge in the wire/well and the difference in
the electronic structure between the various methods, {\it for a given
resultant $n_e$}, is negligible. Thus here we actually vary
a surface gate voltage to change density, but in displaying the results
it is sufficient just to plot variables versus $n_e$.

For large $z$ we find, as noted, that
the solution for the holes is sensitive to the asymptotic
value of the potential. Generally, we assume Neumann boundary
conditions, which is equivalent to assuming complete neutrality
of the substrate. However we discuss below the nature of the 
hole metastable state in the case where the chemical potential is
pinned to a shallow donor level at large $z$.

Since the modulation
doping is $n$-type, the Fermi level is close to the conduction
band throughout the device. Furthermore, since even during 
photoluminescence measurements, the excitation power is very low, 
we ignore the valence band contribution to the charge density.
Therefore, the total density which enters the Poisson equation
is given as
\begin{equation}
\rho(x,z) = N_{+} (x,z) + n(x,z)
\end{equation}
where $n(x,z)$ is the density of conduction band electrons only, and
where the charge density of ionized donors is $N_{+}(x,z)$. 
The conduction band electron density is given in the well
regions by
\begin{equation}
n(x,z) = \int \; dk \; \sum_n f(\mu - k_y^2 - E_n) |\psi_n (x,z)|^2
\label{eq:three}
\end{equation}
where $f$ is the Fermi function, $\mu$ is the chemical potential
(assumed constant throughout the device and taken as the
zero of energy) and $k_y^2$ is the kinetic energy of motion in
the $y$-direction. We also
employ a Thomas-Fermi approximation for the density of
electrons in the barrier regions away from the wells.
Finally, we assume temperature $T = 4 \; K$.

We find little difference in the results whether the donors,
which are silicon atom DX centers \cite{wire}, are placed in
thermal equilibrium with the electron gas or are treated as
a simple sheet of positive charge with fixed areal density.
Insofar as at low temperatures the latter approximation is
more physical, we use this assumption in the calculations
discussed in this paper.

Upon obtaining a self-consistent solution
of Schr\"{o}dinger and Poisson equations for conduction band electrons,
we calculate the hole eigenfunctions with
the potential
\begin{equation}
V_h(x,z) = V_V (x,z) - e \phi(x,z)
\end{equation}
where now $\phi(x,z)$ is the electrostatic potential which
solves the self-consistent problem above. The valence band
offset $V_V (x,z)$ also depends on the aluminum concentration. We
assume a bandgap offset parameter $Q_e = 0.6$. We treat the 
holes in the simplest approximation, assuming a single, heavy-hole
isotropic effective mass which depends only on aluminum
concentration; specifically $m_h^* (GaAs) = 0.377 \; m_0$
and $m_h^* (Al_{0.3} Ga_{0.7} As) = 0.403 \; m_0$ \cite{Eke87}.

To summarize therefore and to make the context of this
calculation clear, we are interested principally in the
BGR as a function of $n_e$, the integral over $x$ and $z$ of
$n(x,z)$, which we imagine to be
modulated experimentally with an electrostatic gate on the
overgrowth surface \cite{Sed99}. Thus we solve the evolving
self-consistent electronic structure assuming the
chemical potential to be far from the valence band (i.e. no holes).
We solve for the hole subbands as a one particle
problem {\it after} solving
the electronic structure at given $n_e$, using the
resultant electrostatic potential. We use a simplified band
structure for the holes, in contrast to ref. \cite{Gol97}
which employs the full, four-band Luttinger Hamiltonian but
which assumes $n_e=0$.

\subsection{Electronic structure}

The parameters for the calculation are summarized in the 
schematic, Fig. 1. We follow reference \cite{Gol97} by
designating the overgrowth quantum well as ``QW2'' and
the initial growth well as ``QW1.'' 

 \begin{figure}[hbt]
 \setlength{\parindent}{0.0in}
  \begin{minipage}{\linewidth}
 \psfig{file=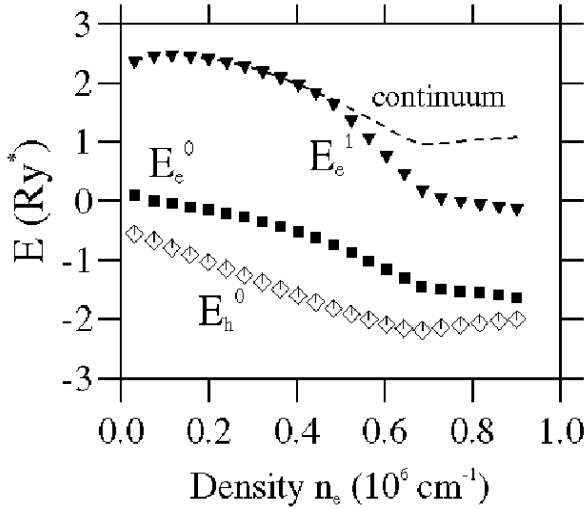,width=8cm,height=7cm}
 \vspace*{3mm}
 \caption{Subband energies vs. wire 1D
density for electron (e) and hole (h).
Electron levels measured with respect to Fermi surface; hole level
with respect to escape to substrate. Electron continuum begins
at dashed line. Filling of $E_e^1$ begins at $n_e \sim 0.5$.}
  \end{minipage}
 \end{figure}

The structure of the subbands of conduction band electrons as
a function of $n_e$ is shown in Fig. 2. The zero of energy for the
electrons is the Fermi energy $E_F$ (the plotted hole energy
is discussed further below).
In the higher density regime
a second subband, which has concentration in the inter-wire QW2
region (cf. Fig. 4), is occupied. The region marked ``continuum'' corresponds
to the beginning of a dense set of states which are asymptotically
free to escape into the substrate (i.e. along QW1
which runs in $+z$ direction). This is so because QW1
is wider than the QW2, and the electrostatic
advantage of proximity to the positive dopants in QW2 is overcome
by the additional confinement energy there. Nonetheless, the
spreading of the {\it occupied} states, when the total density is 
increased, occurs within QW2
(so as to compensate the positive charge of the donors and gate).

It is noteworthy that for 
$n_e \stackrel{>}{\sim} 0.5 \; a_B^{*-1}$ 
($\approx 0.5 \times 10^6 \; cm^{-1}$), there are
two subbands which are below the continuum in contrast to
the situation for lower $n_e$ where only a single bound state
exists. This is also consistent with
the bare ($n_e=0$) TQW studied in Ref. \cite{Gol97}
which had a nearly identical aspect ratio (i.e. ratio of well widths)
and exhibited only a single state separate from the continuum.
The estimate there of a separation between $E_e^0$ and continuum of
$\sim 15 \; meV$ is also consistent with our slightly lower
value of $\sim 13 \; meV$. Interestingly, this 
separation remains approximately
constant (Fig. 2) as $n_e$ is increased, until $E_e^1$ drops down
from the continuum and begins to fill.

 \begin{figure}[hbt]
 \setlength{\parindent}{0.0in}
  \begin{minipage}{\linewidth}
 \psfig{file=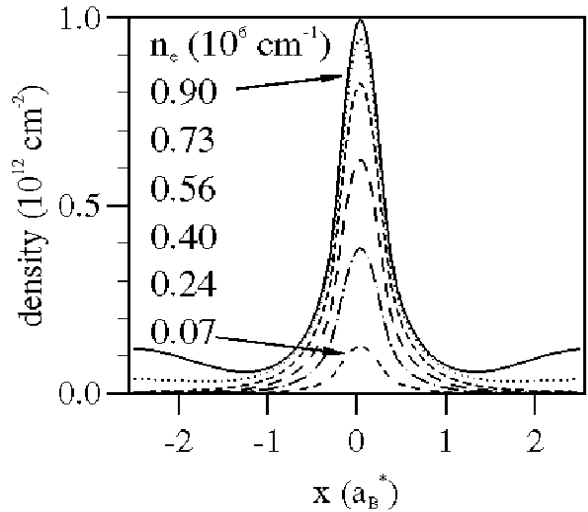,width=8cm,height=7cm}
 \vspace*{3mm}
 \caption{Areal density (integrated along $z$) vs. $x$
for various total wire densities. Confinement to wire almost
complete below $n_e \sim 0.6$. For highest $n_e$ interwire
density develops subsidiary maximum at $\pm 2.5 \; a_B^*$.}
  \end{minipage}
 \end{figure}

In Fig. 3 we have plotted the 2d surface
density as a function of $x$ for various total $n_e$. As $n_e$ 
approaches $1 \; a_B^{*-1}$ some spreading of the density
away from the T-junction occurs. This results both from the occupation
of $e1$ and also from slight wave function spreading of $e0$
(cf. Fig. 4).
Note, however, that,
rather than spreading smoothly out from the wire, the density
achieves a local maximum at an inter-wire minimum in the potential 
(at $\pm 2.5 \; a_B^*$ which,
due to periodicity, are equivalent points),
analogous to the electrostatic potential minima in the barriers of a 
semiconductor superlattice. Here, however, the effect is both electrostatic
and quantum mechanical. Note, in figure 4, where the
moduli squared of the eigenfunctions are plotted, 
that the subband $e1$ has {\it two}
nodes along the $x$-direction (i.e. along QW2). For different
parameters (not shown), such as for a wider QW2 or much higher $n_e$, 
a state with a single node at the T-junction is lower in energy.
But in this regime the attractiveness of the T-junction is 
sufficient to stabilize the even node state. 

 \begin{figure}[hbt]
 \setlength{\parindent}{0.0in}
  \begin{minipage}{\linewidth}
 \psfig{file=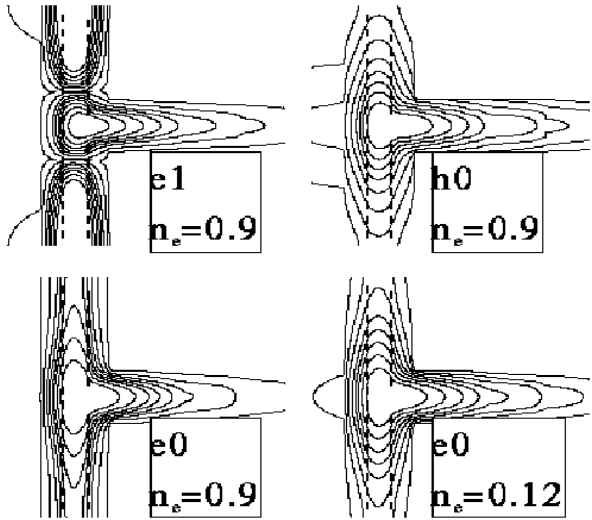,width=8cm,height=7cm,angle=180}
 \vspace*{3mm}
 \caption{Wave functions (moduli squared) showing
spreading of e0 as $n_e$ increases from $0.12 \times 10^6 \; cm^{-1}$
(LR) to $0.9 \times 10^6 \; cm^{-1}$ (LL). First excited electron state
at shows {\it two} nodes along $x$ (UL). Hole ground state (UR)
unchanging and strongly confined at junction.}
  \end{minipage}
 \end{figure}

The spreading of $e0$ with increasing $n_e$ observable in
Fig. 4 is relevant to
the evolution of the gap between conduction and valence
bands. We will see below that the dominant portion of the 
BGR arises from exchange and correlation effects. Nonetheless, 
as noted in the introduction, the difference between the evolution of
electron and hole wavefunctions with $n_e$ produces a kind
of kinematic BGR as in the quantum confined Stark effect.
Explicitly, the hole eigenfunction, which is shown in Fig. 4
only for $n_e = 0.9 \times 10^6 \; cm^{-1}$, undergoes
essentially no change with $n_e$ over the $n_e$ range of Fig. 2. 
Therefore the hole
essentially tracks the electrostatic potential at the T-junction. 
The electrons in $e0$ however, for increasing $n_e$, are able to
lower their energy by spreading along QW2. This implies a
redshift in the bandgap which emerges even in the absence of
exchange-correlation effects (see the Hartree results in Fig. 6
below).

\subsection{Hole states}

The stability of the hole state at the T-junction depends 
weakly on the chosen boundary conditions for Poisson Eq.
at large $z$. If we pin the chemical potential to an assumed
shallow donor level at large $z$, a common assumption, the hole
state at the T-junction is merely metastable, and escape, through
a long, shallow barrier, to the substrate is energetically 
preferred. Assuming Neumann boundary conditions, or ``flat
bands'' at large $z$, on the other hand, results in a true
bound state. Dynamically, the difference between ``pinned band''
and flat band conditions is mostly inconsequential for holes
photo-generated {\it at the junction}. For any reasonable background
(i.e. un-intentional) donor density, the barrier to escape of
the hole is too large and the hole will remain and, presumably,
recombine in the wire. Furthermore, the hole will induce
an image in the electron gas which will cause further, electrostatic
binding to the wire region. A similar effect of holes bound to a 2DEG
has been discussed recently in Ref. \cite{railing}.

For holes generated in the substrate, far from the wire, however,
the dynamics of the diffusion of holes {\it into} the wire region can
be realistically expected to depend on the band shape as reflected
in this dichotomy over boundary conditions. Generally, fewer 
background donors is favorable to hole diffusion from substrate into 
wire.

 \begin{figure}[hbt]
 \setlength{\parindent}{0.0in}
  \begin{minipage}{\linewidth}
 \psfig{file=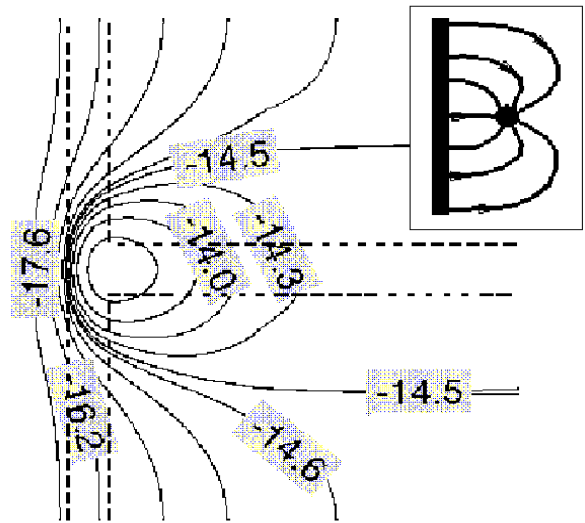,width=8cm,height=7cm}
 \vspace*{3mm}
 \caption{Contour plot of electrostatic potential
for $n_e=0.9 \times 10^6 \; cm^{-1}$. Schematic illustrates field 
lines, which help to localize hole at junction.}
  \end{minipage}
 \end{figure}

Even for pinned band conditions the basic charge distribution
near the wire establishes a purely electrostatic barrier
to hole escape, as seen in Fig. 5.
Here we plot the 2d electrostatic
potential for $n_e = 0.9 \times 10^6 \; cm^{-1}$. 
The potential contour principally derives from a plane of
positive charge (the donor layer) and the line of negative 
charge in the wire. This results in an electrostatic potential
hill near the junction (to which holes are attracted). Thus the hole has
at least a long shallow barrier to escape even before considering
the quantum effect that further binds the hole to the junction.

In Fig. 2 we have plotted the energy of the first hole subband,
$E_h^0$, versus $n_e$. In this case we have used Neumann boundary
conditions for Poisson's Eq. at large $z$ and the energy zero
for $E_h^0$ is the electrostatic potential at large $z$ added
to the confinement energy of QW1; in other words it is
the energy required to escape to the substrate continuum. Here we see
that the state is stable, and not merely metastable. Further,
the increase in the binding with $n_e$ (up to 
$n_e \sim 0.6 \times 10^6 \; cm^{-1}$) arises from the 
strengthening of the electrostatic potential minimum (for holes)
shown in Fig. 5. 

\subsection{Bandgap}

The bandgap in the heterostructure depends
upon the assumed intrinsic bandgap in the host materials, $GaAs$ and
$AlGaAs$ which, as noted, we have taken from the literature
\cite{Zachaw}. The bandgap variation with $n_e$ is obtained from
a sum of the lowest subband energies for the electron and
the hole, each measured with respect to the $GaAs$ band edge at any
common point, then added to the intrinsic bandgap. Figure 6 illustrates
the variation of the bandgap as a function of $n_e$. The rapid change
for small $n_e$ for the full LDA calculation derives from the form
of the exchange-correlation potential. The exchange energy alone, for
comparison, is proportional to the Fermi momentum $k_F = \sqrt{E_F}$
\cite{Mah86}. The overall order of magnitude and shape of the BGR
are consistent with the results of calculations within the GW
approximation \cite{Hwa98}. However, since the authors of 
reference \cite{Hwa98} assume a neutral, two-component plasma, 
the results are not quantitatively comparable.

Also in figure 6 we plot the BGR for the case where the 
exchange-correlation potential is turned off, that is, for pure Hartree.
Here the bandgap still varies, albeit much less, with $n_e$.
The origin of the shift, alluded to in the discussion of
Fig. 4, is the variation of the wavefunction of the electron
with $n_e$ which causes a kinematic component to its
energy while the hole eigenstate is essentially dormant
at the junction. This effect is a slight generalization of
the quantum confined Stark effect, wherein the bandgap
is affected by an electric field, which is varied, 
between the electron location and that of the hole. In our
case the electron additionally shifts its location as
the gate bias varies.

 \begin{figure}[hbt]
 \setlength{\parindent}{0.0in}
  \begin{minipage}{\linewidth}
 \psfig{file=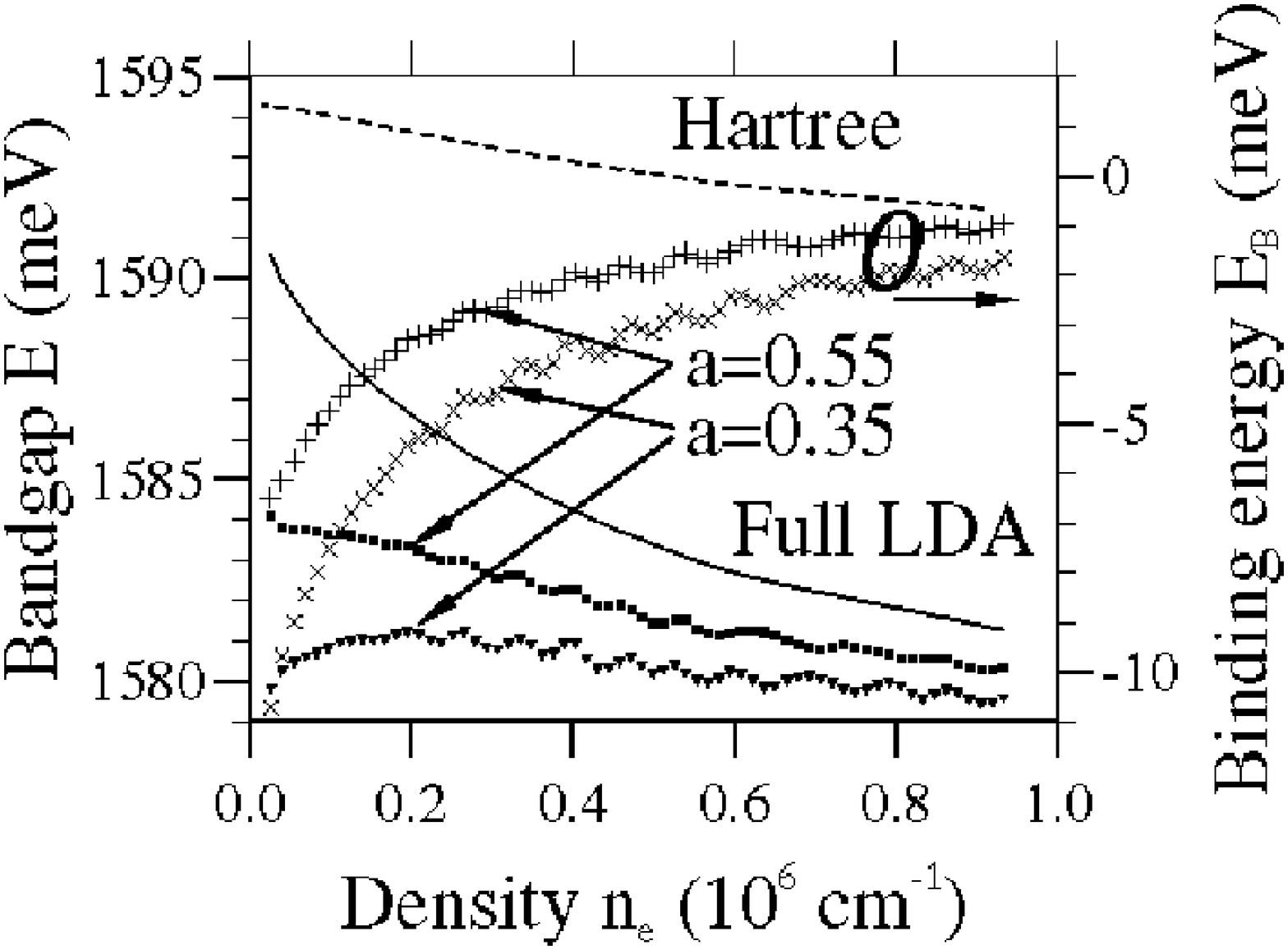,width=8cm,height=7cm}
 \vspace*{3mm}
 \caption{Bandgap renormalization for full LDA (solid) and
for pure Hartree (dashed), here in meV, as a function of density. 
Exciton binding energies for wire radius $a=0.55 \; a_B^*$ (pluses)
and $a=0.35 \; a_B^*$ (crosses) and LDA bandgap corrected for
excitonic energy as squares and triangles, respectively.}
  \end{minipage}
 \end{figure}

We postpone a discussion of the final result of Fig. 6, the
behavior of the BGR when correction for excitonic energy
and screening are included, until after the next section.

\section{Exciton}

The electronic structure of the exciton can be treated in 
a fashion analogous to the calculation of electron binding
to shallow impurities, which was described some time ago 
by Vinter \cite{Vin82} for a silicon inversion layer. In contrast to the
impurity problem though, the hole cannot be treated as a point
charge, but has a wave function which must be computed
self-consistently with those of the bound and free electrons.

\newcommand{\br}{{\bf r}}
\newcommand{\ve}{\varepsilon}
In this section we employ a simplified geometry, shown in
Fig. 7, for a single mode, cylindrical quantum wire along
direction $y$ of
radius $a$ surrounded by a cylindrical metal gate at
radius $R$ whose potential $U_{ext}$ is used to
vary the electron density. The radial charge distribution
of both electrons and holes is assumed to be given by a single
Gaussian wave function 
$\xi_1(r) = \frac{1}{a} \sqrt{\frac{2}{\pi}} e^{-r^2/a^2}$. 
The eigenstates of the electrons
are determined by
\begin{equation}
(-\frac{1}{2m_e} \frac{\partial^2}{\partial y^2} 
+ V_{eff}(y) + \tilde{V}_{xc}(y)) \psi_p(y)
= \ve_e^p \psi_p(y), \label{eq:s1}
\end{equation}
where 
\begin{equation}
V_{eff}(y) \equiv \int d^2 \br \; |\xi_1(r)|^2 e \phi_e(r,y), \label{eq:eff}
\end{equation}
and where the electron effective mass is $m_e$ and $\phi_e(r,y)$
is the electrostatic potential (see below). The length of
the cylinder is $L_y$ and Eq. \ref{eq:s1} is treated with periodic
boundary conditions at the ends $y = \pm L_y/2$. The effective
1D exchange correlation potential at $y$ is determined from
the 3D parameterized form by
\begin{equation}
\tilde{V}_{xc}(y) = 2 \pi \int dr \,r \br \rho_e(r,y) V_{xc} (\rho_e(r,y))
\end{equation}
where $\rho_e(r,y) = |\xi_1(r)|^2 \tilde{\rho}_e (y)$ and where 
$\tilde{\rho}_e (y)$ is determined by filling eigenstates of 
Eq. \ref{eq:s1} up to a fixed chemical potential $\mu$, taken as
the energy zero of the problem. 
 \begin{figure}[hbt]
 \setlength{\parindent}{0.0in}
  \begin{minipage}{\linewidth}
 \psfig{file=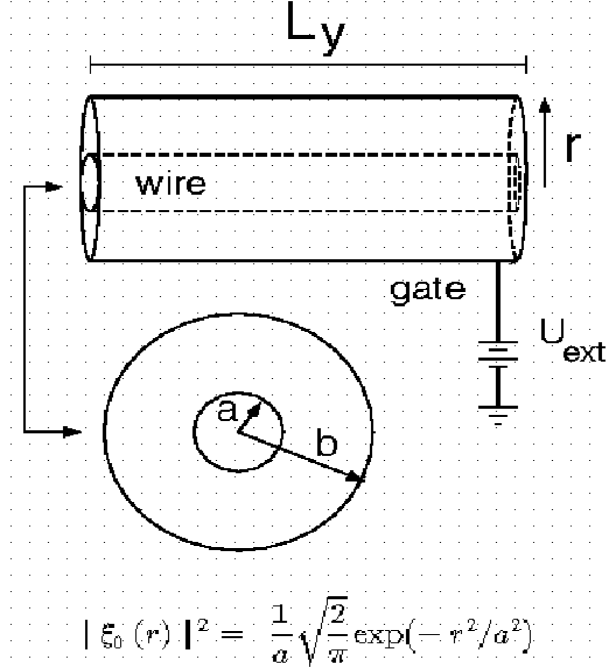,width=8cm,height=9cm}
 \vspace*{3mm}
 \caption{Schematic of cylindrical wire for
model exciton calculation. Hole centered at midpoint of wire.
Radial charge distribution of hole and electron taken as
single Gaussian. Drawing not to scale: $b \stackrel{\sim}{>} 10a$.}
  \end{minipage}
 \end{figure}
The electrostatic potential in
\ref{eq:eff} is determined by solving Poisson's equation
\begin{equation}
-\nabla^2 \phi_e(r,y) = \frac{4 \pi }{\kappa} \rho(r,y)
\end{equation}
subject to Dirichlet boundary conditions at the gate, $r=R$,
and Neumann conditions at $y = \pm L_y/2$. Here $\rho(r,y)$ is
the charge density of both electrons and the single hole.

An expression similar to Eq. \ref{eq:s1}, without 
the exchange-correlation potential, and with the hole mass $m_h$
in place of $m_e$, can be written down for the hole, where only
the lowest eigenstate is filled. Note however that in order to avoid
self-interaction of the hole, the electrostatic potential 
(and the $y$-dependent effective potential) in
this case is calculated from $\rho_e(r,y)$, i.e. the density of
only the conduction band electrons, and is denoted $\phi_h$. 
Self-interaction of the
{\it electrons} is assumed to be compensated for by the exchange-correlation
potential, as is standard in DFT; although it is known that the cancellation
in $LDA$, unlike Hartree-Fock, is not exact. Note that a small inaccuracy
of the model is that, by using, for the hole, the electrostatic potential 
computed without the hole charge $\rho_h(\br)$ included, the image charge
induced by the hole on the gate, with which the hole physically
does interact, is absent. A more precise, though computationally
taxing, approach would be to compute $\phi_h(r,y)$ from all the
charge but then subtract $\int d^3 \br' \rho_h(\br') 
\frac{e}{|\br - \br'|}$, the potential
of the hole in free space.

The binding energy $E_B$ of the exciton is defined as the difference
in the total energy between the case where the hole is localized,
on the one hand, and the case where the single hole charge is spread 
uniformly in $y$, a free hole, on the other hand. The total 
energy must include, in addition to a sum of occupied eigenvalues,
a double-counting correction as well as the energy related to
the gate. Furthermore, this gate energy is modified by the
work supplied to the gate \cite{LP1,Sto95}, and is hence
really a {\it free} energy. The overall expression for 
this free energy is thus
\begin{eqnarray}
F(U_{ext}) & = & \ve_h^1 + \sum_p n_p \ve_e^p \nonumber \\
& - & \frac{1}{2} \int d^3 \br
      (\rho_e(\br) \phi_e(\br) + \rho_h(\br) \phi_h(\br)) \nonumber \\
& + &   \int d^3 \br \rho_e(\br) (\ve_{xc}(\rho_e(\br)) - 
      V_{xc}(\rho_e(\br))) \nonumber \\
& - & \frac{1}{2} Q U_{ext},\label{eq:F}
\end{eqnarray}
where $Q$ is the total charge induced on the gate, determined
by computing the normal derivative of the potential at the gate
surface, and where $n_p$ is the occupancy of state $p$ given
by a Fermi function. Also, the third term on the right hand side
of Eq. \ref{eq:F} is a form of double counting correction for
the exchange-correlation energy, where $\ve_{xc}(\rho)$ is the
exchange-correlation energy per particle of a homogeneous electron
gas of density $\rho$ \cite{DFT}.

We note that one might write the total energy as simply the sum
of occupied Kohn-Sham energies (including the hole) via Koopman's
theorem. To lowest order, where only a single bound
electron and the hole energies are changing appreciably,
one could further estimate this as simply $F(U_{ext}) = \ve_h^1 + \ve_e^1$.
However these are both approximations and the correct formula,
Eq. \ref{eq:F}, must be used for all but heuristic purposes.

\subsection{Results}

In figure 8 we show $E_B$ computed as a function of $U_{ext}$ 
for $a = 0.35, 0.45$ and $0.55 \; a_B^*$. Also plotted is the
 \begin{figure}[hbt]
 \setlength{\parindent}{0.0in}
  \begin{minipage}{\linewidth}
 \psfig{file=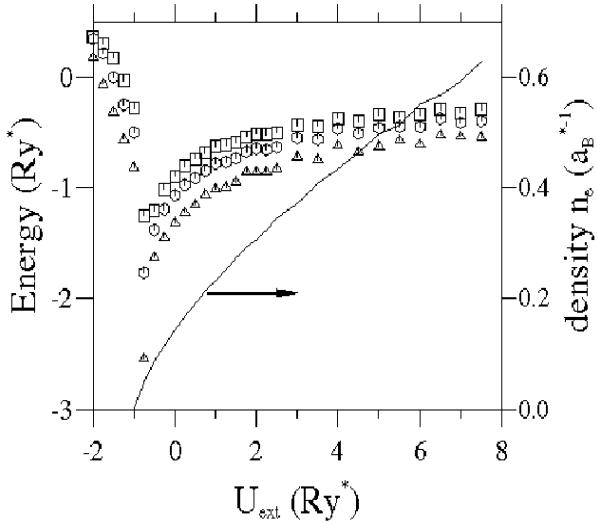,width=8cm,height=7cm}
 \vspace*{3mm}
 \caption{Binding energy of exciton vs. gate
potential for $a=0.35$ (triangles) $0.45$ (hexagons) and $0.55 \; a_B^*$
(boxes). Line gives corresponding $n_e$ (see text for exact definition).
Rise of energy below $n_e=0$ caused by shift of electronic
charge to gate for delocalized hole case.}
  \end{minipage}
 \end{figure}
\noindent conduction band density $n_e$, which is here defined as the
total number of conduction band electrons in the wire for the 
localized hole case,
minus unity (the bound electron) divided by the wire
length $L_y$.
When the gate voltage becomes negative,
all electrons become depleted from the 

 \begin{figure}[hbt]
 \setlength{\parindent}{0.0in}
  \begin{minipage}{\linewidth}
 \psfig{file=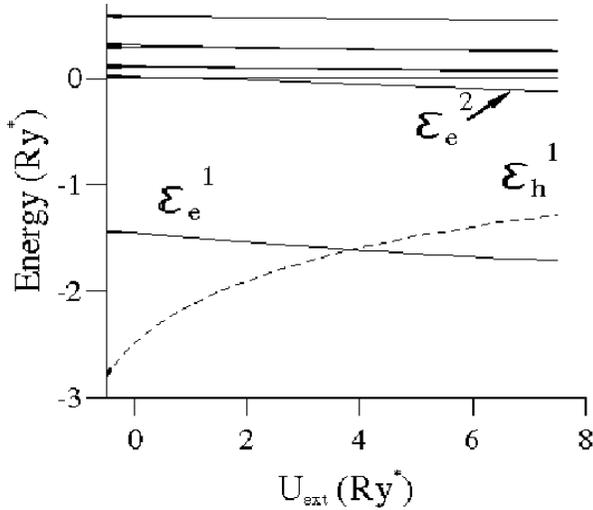,width=8cm,height=7cm}
 \vspace*{3mm}
 \caption{Discrete Kohn-Sham energy levels for
electron (solid) and hole (dashed) vs. gate voltage,
$a=0.35 \; a_B^*$. To
lowest order exciton binding energy is 
$E_B \approx \varepsilon_e^1 + \varepsilon_h^1$, hence 
binding weakens with increased $U_{ext}$ (and $n_e$).
Hole localizes two electrons above $U_{ext} \approx 1 \; Ry^*$.}
  \end{minipage}
 \end{figure}

\noindent wire and, below
$U_{ext} = -1.0 \; Ry^*$, the conduction electrons for 
the delocalized hole case shift from the
wire to the gate. This explains the jump in $E_B$ below
depletion. The minimum of $E_B$ occurs at the lowest density 
before this shift occurs, corresponding to a single electron
in the wire. Clearly the minimum depends somewhat sensitively on
the choice of $a$, and the binding can become quite strong
for very narrow wires, as is well known \cite{San92}.
The small fluctuations visible in the energies result
from the finite size $L_y$ and the discreteness of the
``free'' states.

Of considerable interest is the absence of a binding to
unbinding transition for the exciton.
Even when we extend the density range to $> 2 \; a_B^{*-1}$ 
(i.e. $> 2 \times 10^6 \; cm^{-1}$) and even for the widest
charge distribution, $a=0.55 \; a_B^*$, the ground
state remains bound (not shown). 
In fact, as shown in Fig. 9 (see also Fig. 10), a second electronic
state becomes localized about the hole above $U_{ext} \approx 1.25 \; Ry^*$,
corresponding to $n_e \approx 0.2 \; a_B^{*-1}$, thus forming
a so-called trion \cite{Sch82}. Here we have
plotted the (Kohn-Sham) level energies for the localized
hole case, measured relative to the 
band edge far from the hole, so that an energy above zero
indicates an asymptotically free particle. We find that for even
higher densities, $> 8 \times 10^6 \; cm^{-1}$, a third electron
can become bound to the hole; a
state which we refer to naturally as a ``quadron.''

 \begin{figure}[hbt]
 \setlength{\parindent}{0.0in}
  \begin{minipage}{\linewidth}
 \psfig{file=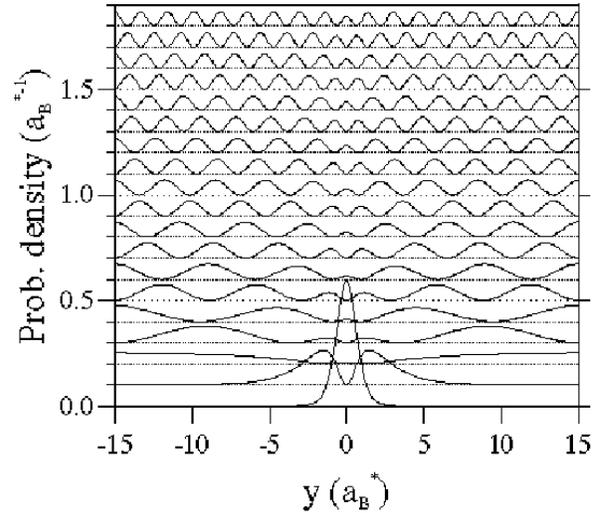,width=8cm,height=7cm}
 \vspace*{3mm}
 \caption{Occupied electronic wave functions
(moduli squared), states $1$ through $19$, $a=0.35 \; a_B^*$,
$n_e = 0.6 \; a_B^{*-1}$. Two states are localized near
hole (at $y=0$). Orthogonality of remaining states to bound
states suppresses screening charge density near hole.}
  \end{minipage}
 \end{figure}

The most intriguing 
feature of the results in Fig. 9 is that, while the hole becomes more
weakly bound, and consequently more spatially extended, the
electrons, particularly the lowest state, become more strongly
bound with increasing $U_{ext}$ (and consequently increasing $n_e$).
A similar result for electron binding to an ionized impurity in
a 2DEG silicon inversion layer was found by Vinter \cite{Vin82}.
This is rather counter-intuitive, since one expects screening, by
free electrons, of the interaction between hole and bound electron
to weaken the attraction and separate the particles. 
Partially this result is understood
as relating to the direct product nature of the exciton state
when expressed in DFT, as opposed to the single composite state
of the two-body or screened two-body problem. Thus the binding energy
is not a single eigenvalue, or even the sum of two eigenvalues,
but rather it must be understood as a difference
between two interacting ground states. 
Nonetheless the decreased energy and increased localization of the
lowest Kohn-Sham level seems puzzling. In figure 10 we show, for
a relatively high density, $n_e=0.6 \; a_B^{*-1}$, the set of
all occupied eigenfunctions (moduli squared). At this $n_e$ the
lowest two electron states are localized near the hole at $y=0$.
Note, however, that due to the restraint of orthogonality, the
densities of all other states in the vicinity of the hole are
{\it suppressed}. Thus while the total conduction band 
 \begin{figure}[hbt]
 \setlength{\parindent}{0.0in}
  \begin{minipage}{\linewidth}
 \psfig{file=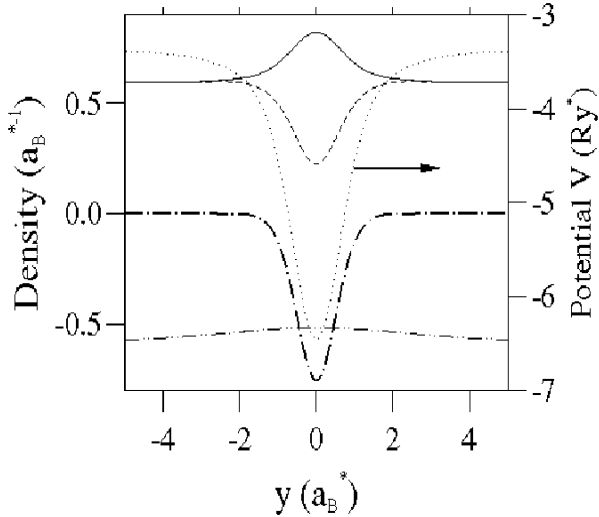,width=8cm,height=7cm}
 \vspace*{3mm}
 \caption{Total electron density (solid), total
density minus bound state (dashed), effective electrostatic
potential $V_{eff}(y)$ (dotted), density of hole (dot-dash)
and density of gate charge (double-dot-dash) versus $y$,
same parameters as Fig. 10.}
  \end{minipage}
 \end{figure}
\noindent density, exhibited in Fig. 11, peaks at $y=0$, this 
density {\it minus}
that of the lowest state diminishes at the hole. Combined with the
uniformly increasing background positive charge on the gate,
this comprises a build-up of positive charge near the hole
as $U_{ext}$ and $n_e$ are increased. Therefore an increased 
density of electrons produces an effectively anti-screening
effect which initially drives the electron closer to the
hole and then admits a second and even a third bound electron. 
The hole eigenstate, meanwhile, is growing increasingly de-localized
and the faster rise of its energy accounts for the increase
of the overall energy (i.e. a decrease of the negative binding
energy). 

\subsection{Combined results}

Finally, in combining the results from Section II and Section III,
we simply add the excitonic binding energy to the bandgap
calculated in LDA for the translationally invariant wire.
The results for this are shown in Fig. 6. Clearly the
exchange-correlation and kinematic effects causing the BGR
produce a red shift which is partially negated by the shift 
of the exciton binding energy due to screening. This is consistent
with the notable insensitivity of the photoluminescence line
position in optical experiments on quantum wires
to either photoexcited electron-hole plasma \cite{Weg93,Amb97}
or to gate-generated 1DEG \cite{Sed99}. Interestingly, it is
also consistent with recent calculations of the dynamically
screened Bethe-Salpeter equation by Das Sarma and Wang
\cite{Das99} for wires with a two-component plasma, although
the interpretation differs somewhat from that discussed here.
While the tendency toward cancellation of BGR and excitonic 
screening is concluded in both studies, 
the most striking difference between the results of Ref. \cite{Das99}
and those here concern the unbinding of the exciton which,
in an electron-hole plasma, is called the Mott transition.
The authors of Ref. \cite{Das99} estimate a merging of the
exciton energy with the electron-hole band edge at
around $n \sim 3 \times 10^5 \; cm^{-1}$, where $n$ is here of course
the density of electrons {\it and} holes. Even in the simple one-electron
static screening approximation they find a vanishing of the exciton
binding energy with density. By contrast, we find that up to the
highest density considered ($3 \times 10^6 \; cm^{-1}$) the exciton
remains bound. Assuming that both conclusions are correct, the
implication is that it is the interaction {\it between} excitons
which leads to their unbinding or, equivalently, their merging
with the continuum.

In any event, the continued existence, despite screening, 
of a bound state of the electron with the hole in one dimension 
is not surprising in that in 1D any small attractive potential
should bind an electron. One could expect that, at some
density the screening of electrons {\it near the Fermi surface}
would effectively cancel the total potential produced by
the hole, thereby releasing the bound electron. However
for this to occur the Fermi wavelength would need
to be much smaller than the exciton radius, which 
is $\sim 1 \; a_B^*$. Our calculations have not, as yet,
shown a density where such a transition of the screening
cloud from states at the bottom of the Fermi sea to states
at the top of the Fermi sea occurs, but should such a transition
exist it would be of considerable interest theoretically.

\section{Conclusion}

In conclusion, we have presented results of density functional 
calculations for the electronic structure of modulation doped
and gated T-shaped quantum wires for the case where the wire can be
assumed to be translationally invariant. We have shown that
the phenomenon of bandgap renormalization can be qualitatively and
quantitatively understood within DFT. We have further employed
a simplified model of a cylindrical quantum wire to examine
the strength of the bound exciton state in the presence of
a one dimensional electron gas. We have normalized out the 
band structure problem here by defining the exciton binding energy
as the difference between the total energy of the wire with
a localized hole and that with the hole charge (and consequently all
electrons) spread uniformly along the wire. We find that 
the variation of the exciton binding energy with density
tends to cancel the bandgap renormalization, in agreement
with recent experiments. Finally, we have noted that, in this
one component plasma case, to the highest densities which we
have considered, no analogue of the Mott transition, i.e.
no unbinding of the electron and hole occurs. Rather, the
orthogonality of the free electron states with those of the
bound electron(s) leads to an anti-screening behavior such that
as $n_e$ increases a second (trion) and even a third (quadron)
bound state forms at the hole.

In the future we hope to investigate the effect of dimensionality
on the exciton physics by extending the calculation to include
multiple 1D subbands.

\begin{center}
{\bf ACKNOWLEDGEMENTS}
\end{center}

I wish to thank Werner Wegscheider, Stefan Sedlmaier, Sankar Das Sarma
and Elisa Molinari
for helpful conversations.

\end{multicols}

\end{document}